\documentclass[fleqn]{niaauth}

\usepackage{graphicx}

\begin{document}
\setlength{\baselineskip}{1.7\baselineskip}

\begin{frontmatter}
\title{Wavelength Shifters for Water Cherenkov Detectors}

\author{Xiongxin~Dai,}
\author{Etienne~Rollin,}
\author{Alain~Bellerive$^{1}$,}
\thanks{Corresponding author. Tel: (613)520-7400 ext. 7537; Fax: (613)520-4061;
e-mail: alainb@physics.carleton.ca}
\author{Cliff~Hargrove,}
\author{David~Sinclair,}
\author{Cathy~Mifflin,}
\and
\author{Feng~Zhang}

\address{Ottawa-Carleton Institute for Physics, Department of physics, Carleton University, 
1125 Colonel By Drive, Ottawa, Ontario K1S 5B6, Canada}

\begin{abstract}
The light yield of a water-based Cherenkov detector can be significantly improved by 
adding a wavelength shifter. Wavelength shifter (WLS) molecules absorb ultraviolet photons and 
re-emit them at longer wavelengths where typical photomultiplier tubes are more sensitive.  In this 
study, several wavelength shifter compounds are tested for possible deployment in the Sudbury 
Neutrino Observatory (SNO).  Test results on optical properties and chemical compatibility for a 
few WLS candidates are reported; together with timing and gain measurements.  A Monte Carlo simulation 
of the SNO detector response is used to estimate the total light gain with WLS.  Finally, a cosmic 
ray Cherenkov detector was built to investigate the optical properties of WLS.

\vspace{1pc}
\end{abstract}


\end{frontmatter}

\section{Introduction}

The Sudbury Neutrino Observatory (SNO) \cite{SNO1} is a heavy water 
Cherenkov detector located at a depth of 2010 m in INCO's Creighton Mine near
Sudbury, Canada. The detector uses 1000 tonnes of ultrapure heavy water as a target
material contained in a 12 m diameter acrylic sphere to detect solar
neutrinos. An array of $\sim$9500 photomultiplier tubes (PMTs), mounted on a
17.8 m diameter stainless-steel geodesic support structure which is immersed
in 7000 tonnes of shielding light water, is used to
observe Cherenkov photons produced in the D$_{2}$O region. 
In recent studies,
the SNO collaboration has provided strong evidence that neutrinos change flavor as they
travel from the Sun to Earth \cite{SNO2,SNO3,SNO4,SNO5,SNO6}, independently of solar
model flux predictions.

Wavelength shifters are generally fluorescent organic chemicals containing
polyaromatic hydrocarbons or heterocycles in their molecules which absorb
photons and re-emit them at longer wavelengths. Previous studies have shown that
adding WLS into Cherenkov detectors increases the amount of detected light by a
factor of 1.2 using amino G \cite{Badino1} and by a factor of 3 using umbelliferone \cite{Willis1}. 

Fig.~1 illustrates why the use of a wavelength shifter is attractive for increasing the light
yield in the SNO detector. The Cherenkov light production rises strongly in the violet end of 
the spectrum but this light is lost due to attenuation in 5.5~cm of acrylic at normal incidence.
The addition of WLS in the D$_{2}$O can boost the Cherenkov signal
without changing the background photons from outside the D$_{2}$O region. This
should significantly increase the detector efficiency and improve the energy resolution of SNO 
which would allow for a better sensitivity of the detector to spectral distortions caused by 
neutrino interactions with matter in the core
of the Sun. In this paper, we studied
the viability to introduce WLS in the SNO detector. Initially, the chemical
and optical properties of several WLS candidates were tested.  Subsequently, a detailed simulation of the 
SNO experiment was performed and a WLS cosmic ray telescope was designed to check the response 
to Cherenkov radiation.
Even though it was subsequently decided that no WLS would be used for a future phase of the SNO 
experiment, we believe the new water based
wavelength shifters that have been found in this study may be useful for some
future Cherenkov detectors and other applications.

\section{Properties of wavelength shifters}
To meet the many requirements for the safe and reliable operation of the SNO detector, a desirable 
WLS should have the following characteristics:
(1) soluble and stable in water;
(2) removable from heavy water as the SNO D$_{2}$O has been loaned by Atomic Energy of
Canada Limited (AECL) and it should be returned additive-free at the end of the
SNO experimental program;
(3) no adverse interaction with the materials used in the detector and water
circulation system, including acrylic, polypropylene, and MnOx beads
\cite{MnOx1} and HTiO absorbent \cite{HTiO1}, which are used for the radium assays
of the heavy water;
(4) high absorbency below 350 nm, and high re-emission probability in the range of 350-500 nm
with a high quantum efficiency in the neutral pH range, with no significant overlap
between the excitation and emission regions; (5) to achieve the best light gain,
the emission spectrum should match, as closely as possible, the optimal PMT sensitive 
region (see Fig.~1);
(6) short fluorescence decay time (a few nanoseconds), as the longer re-emission time
would potentially incorporate trigger ambiguity and could also increase the uncertainty in the 
reconstructed position of the event.

\begin{figure}[ht]
\begin{center}
\includegraphics[width=80mm]{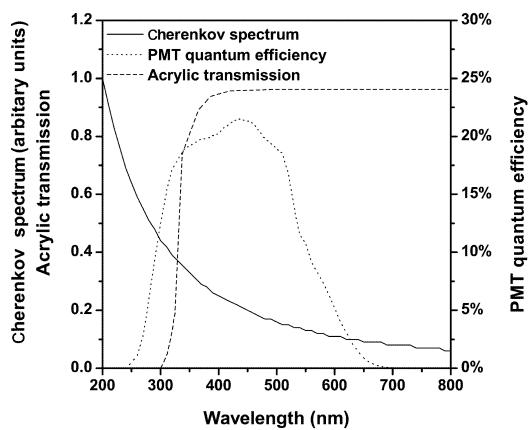}
\caption{Transmission of the SNO acrylic and PMT quantum efficiency as function
of wavelength superimposed on the Cherenkov spectrum (in arbitrary units).}
\label{fig:snorsp}
\end{center}
\end{figure}

Five water-soluble WLS compounds were initially tested according to the above criteria.
They included four coumarin derivatives (7-hydroxy-4-methylcoumarin,
7-hydroxy-4-methylcoumarin-3-acetic acid, 7-hydroxycoumarin-3-carboxylic acid,
and Alexa Fluor 350, all commercially available at Molecular Probes) and
one carbostyril (carbostyril 124, purchased from Sigma-Alderich Co.).  The results are
described as follows.

\subsection{Measurements of optical properties}

The results of the optical measurements for all the wavelength shifters are
summarized in Table 1. Alexa Fluor 350 (AF350) and carbostyril 124 (CS124)
are pH-insensitive in the neutral pH range, whereas the other three coumarin
derivatives show an obvious pH dependence
and their absorption shifts to longer wavelengths as
the pH values increase from 5 to 11. The pH-sensitive coumarins would not be
fully deprotonated, and thus would not be very efficient in converting ultraviolet
photons, until their pH values rise to above 9. The pH value of
the SNO heavy water is 7.  Consequently, a large amount of buffer needs be added
into the detector to alter the water pH if a pH-sensitive WLS is chosen. This is not practical as
it would introduce unnecessary materials and greatly
increase the risk of contamination. 

\begin{table*}
\caption{Summary of optical properties of the wavelength shifters. Typical uncertainties on the 
quantum efficiency and fluorescence lifetime measurements are less then 10\% and 5\%, respectively.}
\protect\label{tbl:1}
\newcommand{\m}{\hphantom{$-$}}
\begin{tabular}{lccccccccccccccccc} \hline
& \multicolumn{5}{c}{7-hydroxy-4-} & Alexa
& \multicolumn{5}{c}{7-hydroxy-4-methyl} & carbostyril
& \multicolumn{5}{c}{7-hydroxycoumarin-} \\ 
& \multicolumn{5}{c}{methylcoumarin} & Fluor 350
& \multicolumn{5}{c}{coumarin-3-acetic acid} & 124
& \multicolumn{5}{c}{3-carboxylic acid} \\ \hline
pH &
5 & 8 & 9 & 10 & 11 & 5-9 &
5 & 8 & 9 & 10 & 11 & 5-9 &
5 & 8 & 9 & 10 & 11 \\ \cline{2-6} \cline{8-12}  \cline{14-18}
Maximum Absorption & & & & & & & & & & & & & & & & & \\
\m $\lambda_{Abs}$ (nm) & 
320 & 336 & 362 & 360 & 360 & 340 &
324 & 326 & 360 & 360 & 360 & 340 &
336 & 386 & 386 & 386 & 386 \\
\m $\epsilon$ (10$^{4}$ L/mol/cm) &
1.4 & 1.1 & 1.6 & 1.8 & 1.8 & 2.0 &
1.5 & 1.3 & 1.7 & 1.8 & 1.8 & 1.6 &
1.0 & 1.1 & 3.2 & 3.3 & 3.3 \\
Maximum Emission & & & & & & & & & & & & & & & & & \\
\m $\lambda_{Em}$ (nm) &
450 & 450 & 447 & 449 & 450 & 443 &
458 & 455 & 455 & 455 & 455 & 417 &
446 & 446 & 446 & 446 & 446 \\
Quantum Efficiency  & 
78\% & 78\% & 87\% & 86\% & 87\% & 92\% &
80\% & 78\% & 82\% & 82\% & 84\% & 97\% &
52\% & 70\% & 89\% & 85\% & 79\% \\
Fluorescence lifetime (ns) &
7.0 & - & - & 6.7 & - & 5.6 &
6.5 & - & - & 6.4 & - & 6.2 &
4.6 & - & - & 4.8 & - \\
\hline \hline
\end{tabular}\\[2pt]
$\lambda_{Abs}$: maximum absorption wavelength; \\
$\epsilon$: molar absorptivity at maximum absorption wavelength; \\
$\lambda_{Em}$: maximum emission wavelength. \\
\end{table*}

\subsubsection{Stability tests of WLS solution}

The stability tests on the WLS solutions were performed by examining the changes
of UV/VIS absorption and fluorescence excitation/emission
over several months of storage in glass vials in the dark.
Two pH-insensitive 1.0 ppm WLS (CS124 and AF350) solutions were tested,
and no significant optical change was observed at neutral
pH condition over a period of more than 6 months.

The solutions of three pH-sensitive coumarins were also tested in UPW at a pH of 5
and in phosphate buffers at a pH of 9. Obvious decreases
were found in the emission intensities of
7-hydroxy-4-methylcoumarin-3-acetic acid both at pH of 5 and 9, and in UV/VIS
absorption intensities of 7-hydroxycoumarin-3-carboxylic acid at a pH of 9 within
two months, indicating a lifetime much shorter than what is required
for the SNO experiment.
Therefore, the pH-sensitive coumarins are
not good candidates for the SNO experiment. In the following sections,
discussion will be mainly focused on the two pH-insensitive compounds AF350 and CS124.
     
\subsubsection{Absorption and fluorescence emission spectra}

The absorption and fluorescence emission spectra were measured using a
Perkin Elmer Lambda 800 ultraviolet and visible (UV/VIS) spectrometer and
a Cary Eclipse fluorescence spectrophotometer made by Varian. Samples were tested 
in 1-cm quartz
cuvettes using ultrapure water (UPW) as a blank. As seen in Fig.~2,
both AF350 and CS124 show strong light absorption below 350 nm and
re-emission in the region of 350-500 nm. At a concentration of 1 $\mu$g/g (ppm),
the attenuation lengths for AF350 and CS124 are respectively 9.3 and 4.8~cm 
at 340 nm. However, CS124 seems to be a better choice for
the SNO experiment as its absorption and re-emission spectra closely match the
response region of the detector. The presence of absorption
at a wavelength higher than 350 nm for AF350 would lead to unnecessary conversion of 
Cherenkov photons in the detectable wavelength range.

\begin{figure}[ht]
\begin{center}
\includegraphics[width=80mm]{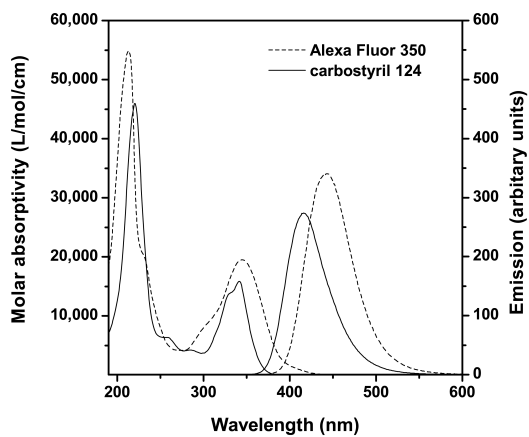}
\caption{The UV/VIS absorption (left) and fluorescence emission spectra (right) for carbostyril 
124 and Alexa Fluor 350.}
\label{fig:spectra}
\end{center}
\end{figure}

\subsubsection{Quantum efficiency measurements}

The quantum efficiency measurements were done by following the Parker-Reas
method \cite{Parker1,Fery1}, which compares the spectral yields of the
WLS solutions to those of standard fluorescence solutions, in our case
quinine sulfate in 0.5 mol/L $H{_2}SO{_4}$ (quantum efficiency 54\% \cite{Quinine1}) 
and harmane in 0.1 mol/L $H{_2}SO{_4}$ (quantum efficiency 83\% \cite{Harmane1}). The quantum 
efficiencies for both AF350 and CS124 are higher than 90\% (see Table~1) with an estimated 
uncertainty of less than 10\%.

\subsubsection{Fluorescence lifetime measurements}
\label{sec:time}
The fluorescence lifetime measurements were carried out using a TimeMaster Laser-based
Fluorescence Lifetime Spectrometer made by Photon Technology International.
As shown in Table~1, the lifetimes are 5.6 ns for AF350 and
6.2 ns for CS124 with an uncertainty of 5\%.  Both are sufficiently
short to meet the reconstruction requirements of the SNO experiment.
 
\subsection{Chemical compatibility and removal tests}

The tests of chemical compatibility of the WLS with acrylic were done by immersing
4$\times$1$\times$0.5 cm$^3$ of acrylic pieces in 25 ml of the WLS solutions and
examining their optical changes on an approximately biweekly basis over
three months. No obvious variations have been seen for any of the five WLS
solutions, suggesting that the loss of WLS on the wall of acrylic
vessel or the interaction of WLS with acrylic should not be an issue in
selection of a candidate.

The impact of the WLS on the two SNO radium assay techniques \cite{MnOx1} \cite{HTiO1} 
was also studied. Small-scale tests were carried out by passing 1 ppm of WLS
solutions through a small plastic column filled with MnOx beads or a small
syringe filter loaded with HTiO absorbent, and the optical properties of
the feed and permeate were measured. All the three pH-sensitive coumarins
show an obvious accumulation and interaction with the MnOx beads and HTiO
absorbent, whereas no change was observed for AF350 or CS124. This indicates
that addition of AF350 or CS124 would not affect the SNO Ra assay techniques.

The Biobeads and activated charcoal used in Milli-Q water systems were tested to extract the
WLS compounds from water. Small scale experiments showed that a reduction factor of about 1000 could
be achievable for the removal of 1.0 ppm WLS from 1000 tonnes of heavy water with activated
charcoal at an equivalent flow rate of 10 l/(min m$^2$) in a single pass. Therefore, it would be
feasible to remove the WLS from the SNO detector after completion of the experiment.

\section{Monte Carlo simulation for SNO}

A detailed Monte Carlo simulation was performed using 
the SNO software (SNOMAN) to account for the 
full light propagation and attenuation, together with 
the complete geometrical acceptance and detector response.  
EGS4 \cite{EGS4} is used in SNOMAN to provide accurate 
propagation of electromagnetic showers.  
The light and heavy water attenuations were set and extrapolated in 
the range of 180-620 nm according to the measurements of Ref. \cite{boivin}. 
The WLS absorption spectrum, quantum efficiency, and wavelength of the 
re-emission peak were parameterized in order to simulate 
the interaction between photons and WLS molecules.  For each 
simulation, SNOMAN interpolated the absorption spectrum 
of Fig.~2 with a fifteen-point piecewise-linear function.  
It also assumed a Gaussian distribution for the re-emission 
wavelength spectrum using the experimentally measured mean 
and width.  According to the measurements of Section 
\ref{sec:time}, the fluorescence lifetimes were assumed 
to be 5.6~ns and 6.2~ns for AF350 and CS124, respectively, 
and the WLS light was emitted isotropically. Furthermore, 
SNOMAN used the WLS concentration to scale the absorption 
coefficient accordingly.

The simulations were performed in order to determine the 
gain of light as a function of the concentration of WLS.  
The concentration values were chosen in the range 
0.01 to 10~ppm based on the mean free path of ultraviolet 
photons in the detector.  A high concentration maximizes 
the number of photons converted and reduces the uncertainty 
in the event position by shortening their mean free path.  
On the other hand, it is important to minimize the quantity 
of WLS for obvious financial reasons and an over-saturation of 
WLS could result in possible self-absorption cycles 
when there is an overlap between the excitation and emission 
regions.

\begin{table}
\begin{center}
\caption{Light gain as a function of the WLS concentration for both Alexa Fluor 350 and 
carbostyril 124.  The results come from a Monte Carlo simulation of the SNO detector.  The 
last column shows the estimated mean free path of Cherenkov light in a WLS solution for 
both candidates.}
\protect\label{tbl:2}
\begin{tabular}{cccc} \\ \hline
 Concentration & Gain  & Gain  & Mean Free Path \\
 (ppm)         & AF350 & CS124 & approx. (cm) \\ \hline
 0.01        & 1.4   & 1.6   & 1000 \\
 0.05	       & 2.0   & 2.4   & 200  \\
 0.10	       & 2.3   & 2.6   & 100  \\
 0.50        & 2.7   & 3.0   & 20   \\
 1.00	       & 2.8   & 3.0   & 10   \\
 5.00	       & 2.9   & 3.0   & 2    \\
10.00	       & 2.9   & 3.1   & 1    \\

\hline \hline
\end{tabular}\\[2pt]
\end{center}
\end{table}

The simulation procedure was straightforward:  10,000 electrons 
with an energy of 10~MeV were generated isotropically 
inside the heavy water.  The gain is defined as 
the ratio of the mean of the number of PMT hits with 
and without WLS.  The gain and the mean free path of Cherenkov light in a WLS solution
are shown in Table~2 for different WLS concentrations.  At 
concentration above 0.50~ppm, there was a saturation 
plateau in the number of detected photons.  Therefore, 
there is no significant advantage in choosing a higher 
concentration in terms of the number of PMT hits.  
However, in order to keep the spatial resolution at a 
reasonable level and bring the absorption mean free path
below 10~cm, the concentration should be 
higher or equal to 1~ppm.  At a few ppm, the light gain 
from the Monte Carlo simulation is estimated to be 2.9 
and 3.0 for the AF350 and CS124, respectively.  

Finally, the energy thresholds with and without WLS were 
identified by simulating SNO data in the pure D$_2$O configuration.
The simulation incorporated both the solar neutrino signal events and all 
internal and external low energy backgrounds. 
In principle, adding a wavelength shifter could allow processes whereby
particles below the Cherenkov threshold  produce light through direct
excitation of the wavelength shifter. In a good scintillator, the
production of light is typically 30,000 photons per MeV of incident
energy. At a concentration of a few ppm, we concluded that this is not
likely to produce more than a few photons and neglected this effect, 
even for the naturally occurring alpha particles of several MeV.
The result showed a clear
shift in the mean number of PMT hit recorded for a typical Cherenkov event. 
At a given statistical significance of the signal over the background
for a 5~MeV analysis threshold without WLS~\cite{SNO3,SNO4}, 
the addition of 1~ppm of WLS would allow for a 3.7~MeV low energy 
threshold.

\section{WLS Cosmic Ray Telescope}

In order to test the WLS compounds in an actual experiment, an apparatus was built that 
uses cosmic rays as the source of Cherenkov light.  
The telescope allows a direct comparison between the 
Cherenkov light produced by cosmic rays and the WLS light.

\subsection{Design of the apparatus}

Two cylindrical barrels made from polyvinylchloride (PVC) 
are placed one above the other allowing a vertical cosmic
ray to go through both, as shown in Fig.~3.
Three scintillator panels placed on the top, the center 
and the bottom of the apparatus are used in coincidence to 
trigger a cosmic ray event.  The top and bottom scintillator 
panels have larger diameters (30 cm) to maximize the 
trigger rate, while the middle one is 
smaller (20 cm) to ensure fully 
contained events. Two layers of lead bricks with a total 
thickness of 20 cm are placed above the bottom scintillator 
panel to eliminate the soft component of the cosmic ray 
shower in the data.  The two barrels are identical in 
size, with a height of 42.5~cm and a diameter of 22.9~cm.  
They both contain a hermetic sample cell that can either be filled 
with pure water or a WLS solution.  These cells are
separated from a Hamamatsu R1408 PMT by an ultraviolet 
transparent (UVT) acrylic window, using the same type of PMT and 
acrylic as in the SNO detector.  The sample cells are 
cylindrical with a diameter of 20.4~cm and a height of 15~cm, 
for a total volume of 4.9 liters of optical medium.

In order to reduce the variation of gain of the PMTs due to their orientation with respect
to the Earth's magnetic field, a cylindrical envelope of mu-metal shields both PMTs
from any variation of the magnetic field. 

Depending on the orientation of the cells (PMT looking upward or downward) and the type of 
optical medium present in each sample cell, it is possible
to isolate the proportion of WLS light and direct Cherenkov light detected by the PMTs and 
perform a light gain measurement. The Cherenkov cone of light always point downward, therefore 
only a PMT in the upward facing position would detect it, while the isotropically distributed WLS 
light is detected by both downward- and upward-looking PMTs.

Although many concentrations of different WLS candidates would
have been possible to analyze with this apparatus, only CS124 at a concentration of 15.4 ppm was 
tested since it seemed the most promising candidate for its addition to the SNO detector.  This 
concentration is higher than the optimal concentration of a few ppm obtained from the Monte Carlo 
analysis for SNO \cite{Rollin} because it was necessary to compensate for the smaller scale of 
the apparatus.
 
\begin{figure}[ht]
\begin{center}
\includegraphics[width=75mm]{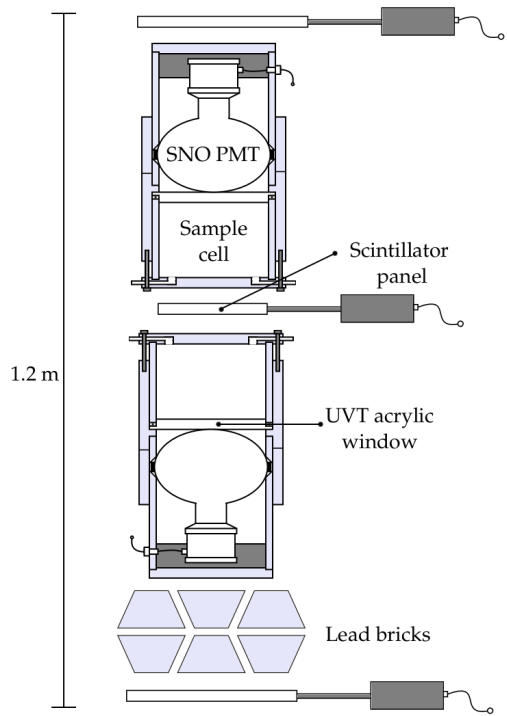}
\caption{Schematic drawing of WLS cosmic ray telescope.}
\label{fig:telescope}
\end{center}
\end{figure}

\subsection{Calibration of the apparatus}
 
Two types of calibration were performed. Since the resolution 
and sensitivity of the SNO PMTs allow the detection of a single photon, 
the first method consists of measuring
the charge induced by a single photon hitting the PMT. Thus, the absolute scale
of the apparatus was measured and resulted in a good linearity up to a few 
thousand photons, which is the luminosity expected from a cosmic ray event.
The second method uses the data itself to check if both PMTs respond the same way to the 
Cherenkov light as a function of the voltage. 
The voltages of the two PMTs were adjusted such as to equalize their gains.
Such calibration allows 
consideration of the intrinsic sensitivity of each PMT and any other cell asymmetry.

\subsection{WLS timing measurements}

The WLS re-emission time measurement consists of doing a fit of the PMTs pulse shape,
assuming two Gaussianly distributed time components, the direct Cherenkov light
and the WLS light.  The fit of the re-emission time of CS124 gave a value of 6.1~$\pm$~0.5~ns, which 
is in good agreement with 6.2~$\pm$~0.3~ns found in the fluorescence lifetime measurement using a 
laser system (see Table~1). Although a more precise measurement would have been possible with
a faster data acquisition system and a more sophisticated pulse shape analysis, it was
unnecessary since the laser system was faster and easier to operate, and much less susceptible
to systematic errors.  Based on the agreement between the timing measurements performed, the WLS 
telescope seems to be well calibrated for the gain measurement described in the next section.

\subsection{WLS light gain measurement}

Depending on the orientation of the Cherenkov cell, there are two ways to measure the mean 
amplitude of pulses with and without WLS.  If a WLS solution is in the upper cell and water is 
in the lower cell, only WLS light is detected in the first ($A_W$) and only Cherenkov light 
is detected in the second ($A_C$).  
After inverting the configuration, both Cherenkov and WLS light 
are detected in the lower cell ($A_{W+C}$), while in the upper cell only the dark noise 
should be detected ($A_{\rm{noise}}$). Based on the 
ratio $\frac{A_{W+C}}{A_C}$ between the mean pulse amplitudes of the PMTs, an increase of 2.0~$\pm$~0.2 in the 
number of detected photons was obtained with WLS.  The error is dominated by the level of noise 
in the PMTs and some residual asymmetry in the cells. A fully consistent results is obtained
when the gain is computed from $\frac{A_W+A_C}{A_C}$.

The main interest of the WLS cosmic ray telescope was to perform a light gain measurement and 
to ultimately determine not only the number of photons detected by the PMTs but the total 
number of photons produced.
Since the PMT in a given cell detects only a fraction of the light produced and this fraction is 
different for Cherenkov and WLS light, a calculation is required to obtain the increase of 
photon produced within a $4\pi$ solid angle. 
The Monte Carlo method was used for the determination of the WLS  cosmic ray telescope acceptance 
and the light gain correction factor.
The simulation generated Cherenkov photons, calculated their mean free path in the WLS solution 
and if the interaction point was within the cell volume, it changed the direction of the photon 
randomly.  It has been found that 61.0\% of the Cherenkov photons propagating toward the PMT 
were hitting it, while only 18.1\% of the isotropic WLS photons ended their path on the PMT.  Therefore, 
adding 15.4 ppm of carbostyril 124 into the WLS cosmic ray telescope 
resulted in a net light gain of of 4.4~$\pm$~0.5.

\section{Conclusions}

The optical and chemical properties of two new water based wavelength shifter (WLS) molecules,
carbostyril 124 (CS124) and Alexa Fluor 350 (AF350), were tested as candidates to increase the 
detection efficiency of Cherenkov light 
of the Sudbury Neutrino Observatory (SNO). The tests indicated that these
pH-insensitive WLS chemicals have strong absorbency below 350 nm and high re-emission probability
between 350-500 nm with a short fluorescence decay time. This can significantly boost
the Cherenkov signals detected by the SNO PMTs which surround the D$_{2}$O region. Their long
lifetime stability and good chemical compatibility with the materials used in the
SNO detector as well as in the heavy water system allow addition of WLS with no major
modification to the detector. Monte Carlo simulations allowed a detailed study 
of the response of a full-scale solar neutrino
experiment to the addition of WLS.  It was shown that a 5 ppm concentration of AF350 or CS124 would 
increase the light yield detected by the SNO PMTs by a factor of 2.9 and 3.0, respectively.  A cosmic 
ray telescope was also built to test the WLS compounds in a well controlled experimental setup, and
an increase of the detected Cherenkov light has been found by adding carbostyril 124. Taking into 
account the geometrical acceptance of the detector, an increase of 4.4~$\pm$~0.5 light yield has 
been measured. The measured increase in number of photons obtained with the WLS telescope has to be 
compared with the light gain of 3.0 obtained with the full simulation of SNO.
It is important to note that 
a Cherenkov detector with a low sensitivity to ultraviolet light will greatly benefit from the addition 
of WLS. While a larger light gain is obtained in a small scale detector, where the attenuation is 
negligible, a reduced increase in the number of detected Cherenkov photons is expected if a wavelength 
shifter would be added in the heavy water of the Sudbury Neutrino Observatory. 
Ultimately the properties 
of the wavelength shifter candidates surveyed in this study can be used toward the conceptual design 
of future water based Cherenkov detectors.

\section{Acknowledgments}
The authors would like to thank Alex Davis, Pascal Elahi, Rachel Faust, and
Christian Mallet for their contributions to the experiments.
We wish to thank Mark Chen and Alex Wright for the measurements
of fluorescence decay time of the WLS chemicals using the laser system in the Department of Physics
at Queen's University.
This research was supported in Canada by 
the Canadian Foundation for Innovation (CFI), the Canada Research Chair (CRC) program, and
the Natural Sciences and Engineering
Research Council (NSERC).

\end{document}